\documentclass[pra,twocolumn,showpacs]{revtex4}
\usepackage{amsmath,amssymb,mathrsfs}
\usepackage{graphicx}
\usepackage{amsmath}
\usepackage{epsfig}
\usepackage[usenames]{color}
\newcommand{\beq}{\begin{equation}}
\newcommand{\eeq}{\end{equation}}
\newcommand{\beqa}{\begin{eqnarray}}
\newcommand{\eeqa}{\end{eqnarray}}
\newcommand{\ba}{\begin{array}}
\newcommand{\ea}{\end{array}}

\begin{document}

\title{Two-dimensional quasi-ideal Fermi gas with Rashba spin-orbit coupling}

\author{Luca Salasnich}

\affiliation{Dipartimento di Fisica e Astronomia ``Galileo Galilei'' 
and CNISM, Universit\`a di Padova, Via Marzolo 8, 35131 Padova, Italy}

\begin{abstract}
We investigate the zero-temperature properties of a 
quasi-ideal Fermi gas with Rashba spin-orbit coupling. 
We find that the spin-orbit term strongly affects the speeds 
of zero sound and first sound in the Fermi gas, 
due to the presence of a third-order quantum phase transition. In addition, 
including a 2D harmonic confinement we show that also the 
shape of the density profile of the cloud crucially 
depends on the strength of the Rashba coupling.  
\end{abstract}


\maketitle

Artificial spin-orbit coupling 
has been recently implemented in bosonic \cite{so-bose,so-bose2} 
and fermionic \cite{so-fermi1,so-fermi2} atomic gases, by
means of counterpropagating laser beams which couple two internal hyperfine
states of the atom by a stimulated two-photon Raman transition.  
These experimental achievements have triggered several 
theoretical investigations in understanding the spin-orbit effects
with Rashba \cite{rashba} and Dresselhaus \cite{dress} terms 
in Bose-Einstein condensates \cite{stringa1,stringa2,burrello,
so-solitons,so-bright1,so-bright2,so-bright3,so-vortices,so-dipolar} 
and also in the BCS-BEC crossover of superfluid fermions 
\cite{shenoy,shenoy2,gong,hu,yu,iskin,yi,dms,iskin2,zhou,jiang,sa,chen,
zhou2,yangwan,iskin3,sa2,he,liu,ne1,ne2}. 

In this Brief Report we focus on the effects of the Rashba spin-orbit 
coupling in a normal two-spin-component (i.e. two atomic species) 
Fermi gas and, for the sake of simplicity and clarity, 
we consider a two-dimensional (2D) quasi-ideal 
atomic gas, for which the inter-particle interaction can be neglected 
in the equation of state. Under the conditions 
of reduced dimensionality and suppressed inter-particle 
interaction (see also \cite{sala-fermi-reduce}), 
we analyze theoretically statical and dynamical properties 
of the system. In particular, after deriving the 
single-particle density of states of our 2D Fermi gas, we determine 
the zero-temperature equation of state, namely the chemical 
potential as a function of the total density and of the Rashba strength. 
From the equation of state we calculate the speed of sound 
both in the collisionless regime (zero sound) and collisional regime 
(first sound). Then we switch on a trapping harmonic potential 
in the 2D Fermi gas showing how the density profile of the fermionic 
cloud depends on the Rashba coupling. 

The two-spin-component single-particle 
quantum Hamiltonian ${\hat h}_{sp}$ of a confined 
two-component 3D Fermi gas of identical atoms with mass $m$ 
and Rashba spin-orbit coupling reads
\beq 
{\hat h}_{sp} = \left( {{\hat p}^2\over 2m} + U({\bf r}) \right) \, \sigma_0 
+ v_R \ 
\left( \sigma_x {\hat p}_y - \sigma_y {\hat p}_x \right) \; , 
\eeq
where $U({\bf r})$ is the trapping potential, ${\hat p}^2=-\hbar^2 \nabla^2$, 
${\hat p}_x=-i\hbar {\partial\over\partial x}$, 
${\hat p}_y=-i\hbar {\partial\over\partial y}$, 
$v_R$ is the Rashba couping constant (Rashba velocity), and 
$$
{\hat \sigma_0} = \left( 
\begin{array}{cc}
1 & 0 \\
0 & 1 
\end{array}
\right) \; , 
\quad\quad 
\sigma_x = \left( 
\begin{array}{cc}
0 & 1 \\
1 & 0 
\end{array}
\right) \; , 
\quad\quad
\sigma_y = \left(
\begin{array}{cc}
0  & -i \\
i  & 0
\end{array}
\right) \;  
$$
are Pauli matrices. We suppose that the external trapping 
potential is given by 
\beq 
U({\bf r})= {1\over 2} m \left[ \omega_{\bot}^2(x^2+y^2)+ \omega_z^2z^2 
\right] \; ,  
\label{potential}
\eeq
where $\omega_z\gg \omega_{\bot}$ (disk-shaped configuration) 
and moreover $\hbar \omega_z/2 \gg \bar{\mu}-\hbar\omega_z/2$ with $\bar{\mu}$ 
the 3D chemical potential of the system. Under these conditions 
the Fermi system is two-dimensional \cite{sala-fermi-reduce}. 

We start our investigation of the 2D Fermi gas of atoms with Rashba 
spin-orbit coupling setting $\omega_{\bot}=0$, which corresponds to a
uniform configuration in the $(x,y)$ plane within a square of area $L^2$. 
In this case, the solution $\phi_{{\bf k},j}({\bf r})$ of the single-particle 
Schr\"odinger equation 
\beq 
{\hat h}_{sp} \, \psi_{{\bf k},j}({\bf r}) = \epsilon_{{\bf k},j} 
\, \psi_{{\bf k},j}({\bf r}) \; 
\eeq
where ${\bf k}=(k_x,k_y)$ is the planar wavevector and $j=-1,1$ is the 
helicity index, has the form \cite{sala-fermi-reduce,lipparini}
\beq 
\psi_{{\bf k},j}({\bf r}) = { e^{i(k_xx+k_yy)-z^2/(2a_z^2)} \over L 
\pi^{1/4} a_z^{1/2} } \, \left( 
\begin{array}{cc}
{1\over \sqrt{2}}  \\
- j \, {i\over \sqrt{2}} \, e^{i\arctan{(k_y/k_x)}}  
\end{array}
\right) \; 
\eeq
with $L$ the characteristic length of the planar 
wavefunction, $a_z=\sqrt{\hbar/(m\omega_z)}$ the characteristic 
length of the axial Gaussian wavefunction, and clearly $i=\sqrt{-1}$. 
The corresponding single-particle 
energy $\epsilon_{{\bf k},j}$ of 2D fermionic particles 
with Rashba spin-orbit coupling is given by 
\beq 
\epsilon_{{\bf k},j} = {\hbar^2k^2\over 2m} + j \hbar v_R 
\sqrt{k_x^2+k_y^2} + {1\over 2} \hbar \omega_z \; . 
\eeq

The 2D single-particle density of states  $\rho(\epsilon)$ 
of non-interacting particles 
with spin-orbit coupling is defined as 
\beq 
\rho(\epsilon) = {1\over L^2} \sum_{\bf k} 
\sum_{j=-1,1} 
\delta(\epsilon_{{\bf k},j}-(\epsilon +\hbar\omega_z)) \; ,  
\eeq
with $L^2$ the area of the system in the plane $(x,y)$. Notice that 
we have shifted the single-particle energy $\epsilon$ to remove the constant 
axial energy $\hbar\omega_z$. 
This single-particle density 
of states $\rho(\epsilon)$ 
can be then easily calculated in the two-dimensional continuum, where  
\beq 
\rho(\epsilon) = \int {d^2{\bf k}\over (2\pi)^2} \ 
\sum_{j=-1,1} 
\delta\left({\hbar^2k^2\over 2m}+j \hbar v_R 
\sqrt{k_x^2+k_y^2} - \epsilon \right) \; .   
\label{pp3}
\eeq
In the case of a 3D uniform gas an analytical formula for $\rho(\epsilon)$ 
has been found by Hu and Liu \cite{huliu}. For our 2D problem we obtain 
\beq 
\rho(\epsilon) = {m\over \pi\hbar^2} 
\left\{ 
\begin{array}{ccc}
0 & \mbox{for} & \epsilon < -\epsilon_R \\
{\sqrt{\epsilon_R} \over \sqrt{\epsilon+\epsilon_R}} & 
\mbox{for} & -\epsilon_R \leq \epsilon < 0 \\ 
1 & 
\mbox{for} & \epsilon \geq 0 
\end{array} \right. \; , 
\label{rho-sp-2d}
\eeq 
with $\epsilon_R=m v_R^2/2$ the characteristic energy of the 
Rashba spin-orbit coupling. We stress that $\rho(\epsilon)$
becomes the familiar 2D constant density of 
states $\rho(\epsilon)=m/(\pi\hbar^2)$ when $v_R=\epsilon_R=0$. 

At zero temperature, the 2D number density $n$ of the ideal 
2D Fermi gas with spin-orbit coupling is defined as  
\beq 
n = \int_{-\infty}^{+\infty} d\epsilon \, \rho(\epsilon)
\, \Theta(\mu - (\epsilon + \epsilon_R) ) \; ,  
\eeq
where $\Theta(x)$ is the Heaviside step function, which is the 
zero-temperature limit of the Fermi-Dirac distribution, and $\mu$ 
is the zero-temperature 2D chemical potential of the system. 
Notice that we have shifted the single-particle energy $\epsilon$ to 
get a non-negative 2D chemical potential $\mu$, moreover 
the 3D chemical potential $\bar{\mu}$ of the system is related to the 2D 
chemical potential $\mu$ by the simple expression $\bar{\mu}
=\mu+\hbar\omega_z/2$. By using Eq. (\ref{rho-sp-2d}) we find 
\beq 
n = {m\over \pi\hbar^2} 
\left\{ 
\begin{array}{ccc}
2\sqrt{\epsilon_R \, \mu} & 
\mbox{for} & 0\leq \mu < \epsilon_R \\ 
\mu + \epsilon_R & 
\mbox{for} & \mu \geq \epsilon_R 
\end{array} \right. \; . 
\eeq
It is easy to invert this formula obtaining the 2D chemical 
potential $\mu$ as a function of the 2D number density $n$, namely 
\beq 
{\mu\over \epsilon_R} = 
\left\{ 
\begin{array}{ccc}
{n^2\over n_R^2} & \mbox{for} & 0 \leq n < n_R \\
-1 + 2 {n\over n_R} & 
\mbox{for} & n \geq n_R 
\end{array} \right. \; ,  
\label{echem}
\eeq
where 
\beq 
n_R={m^2v_R^2\over \pi\hbar^2}
\eeq
is the 2D critical Rashba density. 
It is important to observe that in the absence of spin-orbit coupling, 
i.e. for $v_R=\epsilon_R=n_R=0$, the chemical potential $\mu$ 
becomes the familiar Fermi energy $\epsilon_F=\pi\hbar^2n/m$ of the 
2D ideal Fermi gas and the corresponding Fermi velocity reads 
$v_F=\sqrt{2\epsilon_F/m}$. At $n=n_R$ one finds that 
${\partial^2\mu\over \partial^2n}$ has a jump, and this implies 
a third-order phase transition \cite{kerson}. It is a ``quantum'' 
phase transition because the phase transition 
holds at zero temperature. 

\begin{figure}[tbp]
\begin{center}
{\includegraphics[width=8.cm,clip]{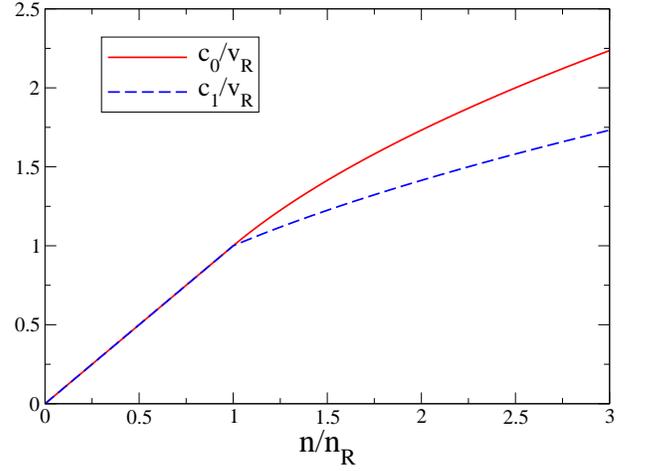}}
\end{center}
\caption{(Color online). Zero-sound and first-sound velocities of the uniform 
2D quasi-ideal Fermi gas with Rashba spin-orbit coupling. 
Solid line: scaled zero-sound velocity 
$c_0/v_R$ as a function of the scaled number density $n/n_R$. 
Dashed line: scaled first-sound velocity 
$c_0/v_R$ as a function of the scaled number density $n/n_R$. 
Here $\epsilon_R=mv_R^2/2$ is the Rashba energy 
and $n_R=2m\epsilon_R/(\pi \hbar^2)$ is the critical Rashba density, 
with $v_R$ the Rashba coupling (Rashba velocity).} 
\label{fig1}
\end{figure}

Up to now we have analyzed a two-spin-component 
2D Fermi gas with spin-orbit coupling, but we have neglected 
the effect of the s-wave scattering length $a_s$ between atoms 
in the equation of state. This assumption of 
``quasi-ideal gas with spin-orbit'' is reliable under the condition 
\beq 
g_{2D} \, n \ll \mu \; , 
\label{quasideal}
\eeq
where $g_{2D}$ is the 2D effective interaction strength 
given by $4\pi\hbar^2a_s/(ma_z\sqrt{2\pi})$ with $a_z$ the 
characteristic length of axial harmonic confinement, 
while the chemical potential $\mu$ is given by Eq. (\ref{echem}). 
Thus, if Eq. (\ref{quasideal}) holds the effect of interaction can be 
neglected in the zero-temperature equation of state $\mu=\mu(n)$. 
Nevertheless, also under the condition (\ref{quasideal}) 
the s-wave scattering length $a_s$ is very important because it 
determines the collisional time $\tau_c$ of the system, 
given by 
\beq 
\tau_{c} = {1\over n_{3D} \, \sigma \, v_F} \; , 
\eeq
where $n_{3D}=n/(a_z \sqrt{2\pi})$ is the 3D number density, 
$\sigma=4\pi a_s^2$ is the scattering length and $v_F$ is the 2D 
Fermi velocity, a density wave propagates with a dispersion relation 
\beq 
\omega = c_s \, q 
\eeq
with $\omega$ the frequency of oscillation, $q$ the wavenumber 
and $c_s$ the speed of sound. 

In the collisionless regime, where 
$\omega \tau_c \gg 1$, the sound is called zero sound and the 
corresponding zero-sound velocity $c_s=c_0$ is given by 
the Landau formula \cite{sala-fermi-reduce,lipparini}
\beq 
c_0 = \sqrt{2\mu\over m} \; .   
\eeq
By using Eq. (\ref{echem}) we immediately find 
\beq 
{c_0\over v_R} = 
\left\{ 
\begin{array}{ccc}
{n\over n_R} & \mbox{for} & 0 \leq n < n_R \\
\sqrt{-1+2{n\over n_R}} & 
\mbox{for} & n \geq n_R 
\end{array} \right. \; ,  
\label{c0}
\eeq
where $v_R=\sqrt{2\epsilon_R/m}$ is the Rashba velocity with 
$\epsilon_R=m v_R^2/2=\pi \hbar^2n_R/(2m)$ 
the Rashba energy. Notice that, according to our definitions, 
$v_R/v_F=\sqrt{n_R/(2n)}$. Moreover, in the absence of spin-orbit coupling, 
i.e. for $v_R=\epsilon_R=n_R=0$, the 2D zero-sound velocity $c_0$ 
becomes the 2D zero-sound velocity of the 2D ideal Fermi gas, 
which is nothing else than the 2D Fermi 
velocity $v_F=\sqrt{2\pi\hbar^2n/m^2}$. 

In the collisional regime, where 
$\omega \tau_c \ll 1$, the sound is called first sound and the 
corresponding first-sound velocity $c_s=c_1$ is given 
by the thermodynamics formula \cite{sala-fermi-reduce,lipparini}
\beq 
c_1 = \sqrt{{n\over m}{\partial \mu\over \partial n}} \; . 
\label{forseno}
\eeq
By using Eq. (\ref{echem}) we immediately find  
\beq 
{c_1\over v_R} = 
\left\{ 
\begin{array}{ccc}
{n\over n_R} & \mbox{for} & 0 \leq n < n_R \\
\sqrt{n\over n_R} & 
\mbox{for} & n \geq n_R 
\end{array} \right. \; .  
\label{c1}
\eeq
Notice that, in general, for a uniform
superfluid the first sound velocity can be obtained either 
by Eq. (\ref{forseno}) 
or from a detailed calculation of vertex function (i.e. within 
Random Phase Approximation). These two may be considered 
as macroscopic and microscopic sound velocity, respectively. 
In the spin-orbit coupled system, the Galilean 
invariance is broken and the macroscopic and microscopic 
sound velocities could be different: we are presently working on 
this puzzling issue. 
From Eq. (\ref{c1}) one finds that, in the absence of spin-orbit coupling, 
i.e. for $v_R=\epsilon_R=n_R=0$, the 2D first-sound velocity $c_1$ 
becomes the 2D first-sound velocity of the 2D ideal Fermi gas, 
namely $c_1=v_F/\sqrt{2}$ with $v_F=\sqrt{2\pi\hbar^2n/m^2}$ 
the 2D Fermi velocity of the ideal Fermi gas. 

In Fig. 1 we report both zero and first sound velocities 
as a function of the scaled number density $n/n_R$. For $n<n_R$ 
the two velocities coincide while for $n>n_R$ they have a different 
behavior. Without spin-orbit coupling, i.e. for $v_R=\epsilon_R=n_R=0$, 
it is immediate to find that $c_0=v_F$ and $c_1=v_F/\sqrt{2}$, which are 
the text-book results of a 2D quasi-ideal Fermi gas \cite{lipparini}. 

We now switch on the soft harmonic potential in the 
$(x,y)$ plane, i.e. we consider the case $\omega_{\bot}\neq 0$ 
in Eq. (\ref{potential}). Within the local density (Thomas-Fermi) 
approximation \cite{lipparini} 
we perform the following shift in the 2D chemical potential 
\beq 
\mu \to \mu - {1\over 2} m \omega_{\bot} r^2 \; , 
\label{shift}
\eeq
where $r=\sqrt{x^2+y^2}$ is the cylindric radial coordinate. 
By using Eq. (\ref{shift}) into Eq. (\ref{echem}) we obtain 
the 2D local number density $n(r)$, which is the density profile 
of fermionic cloud. 

\begin{figure}[tbp]
\begin{center}
{\includegraphics[width=8.cm,clip]{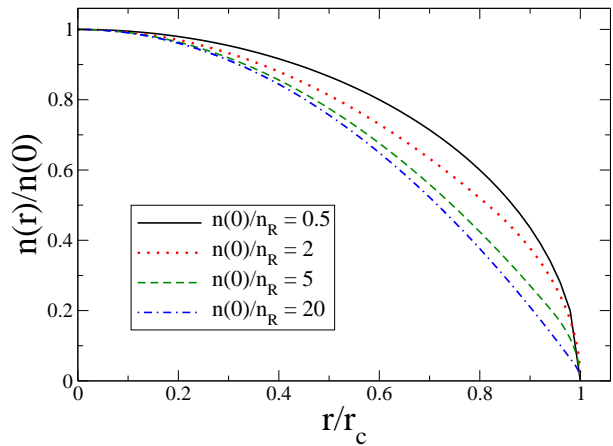}}
\end{center}
\caption{(Color online). Radial density profile $n(r)$ of the uniform 
2D quasi-ideal Fermi gas with Rashba spin-orbit coupling in a harmonic trap. 
The four curves correspond to different values of the ratio $n(0)/n_R$ 
where $n(0)$ is the 2D local density at the center of the cloud 
and $n_R=m^2v_R^2/(\pi \hbar^2)$ is the characteristic Rashba density, 
with $v_R$ the Rashba coupling (Rashba velocity).} 
\label{fig2}
\end{figure}

It is not difficult to show that 
the density profile crucially depends on its value $n(0)$
at the center of the harmonic trap. In particular, if $n(0)<n_R$ 
we find 
\beq 
n(r) = n(0) \sqrt{1-{r^2\over r_c^2}} \quad \mbox{for} \quad 
0\leq r \leq r_c \; , 
\eeq
where $r_c=r_R (n(0)/n_R)$ is the critical radius 
such that $n(r_c)=0$, and $r_R=\sqrt{2\epsilon_R/(m\omega_{\bot})}$ 
is the Rashba radius with $\epsilon_R=\pi\hbar^2n_R/(2m)$. 
Obviously, in this case $r_c<r_R$. Instead, if $n(0)>n_R$ we obtain 
\beq 
n(r) = 
\left\{ 
\begin{array}{ccc}
n(0) \big( 1-{r^2\over r_b^2} \big) & \mbox{for} & 0 \leq r < r_0 \\
n(0) {1-{r_0^2\over r_1^2}\over \sqrt{1-{r_0^2\over r_c^2}}} 
\sqrt{1-{r^2\over r_c^2}}
& \mbox{for} &  r_0 \leq r \leq r_c
\end{array} \right. \; , 
\eeq
where $r_0=r_R \sqrt{2(n(0)/n_R)-2}$, $r_1=r_R \sqrt{2(n(0)/n_R)}$, 
and $r_c=r_R \sqrt{2(n(0)/n_R)-1}$. 
Again $r_R=\sqrt{2\epsilon_R/(m\omega_{\bot})}$ 
is the Rashba radius and moreover $r_0< r_R < r_c<r_1$. 
In Fig. 2 we report the radial density profile $n(r)$ 
for four diffeent values of the ratio $n(0)/n_R$. 
We actually plot the scaled radial density $n(r)/n(0)$ as a function 
of the scaled cylindric radius $r/r_c$, such that both $n(r)/n(0)$ and 
$r/r_c$ are confined in the interval $[0,1]$. Notice that the solid curve 
obtained with $n(0)/n_R=0.5$ is indeed the same for any ratio 
$n(0)/n_R$ between $0$ and $1$. Instead for $n(0)/n_R>1$ our scaled 
density profile depends on the chosen ratio, as shown in the figure. 

In conclusion, we have shown that the inclusion of a Rashba 
spin-orbit coupling in a quasi-ideal 2D Fermi gas 
implies the existence of a critical Rashba density, which depends 
on the Rashba coupling strength, at which the equation of state 
and other physical properties, e.g. speed of sound and density 
profiles, drastically change. Indeed at this critical Rashba 
density there is a third-order phase transition, 
because, as shown by Eq. (\ref{echem}), the second derivative of the 
chemical potential $\mu$ with respect to the number density $n$ 
has a jump at the critical Rashba density $n_R$. 
We believe our predictions can be 
experimentally tested with the available setups. 

\section*{Acknowledgments}

The author thanks Gauri Shankar Singh, Flavio Toigo, and Antonio Trovato 
for useful discussions. The author acknowledges 
Universit\`a di Padova, Cariparo Foundation, and 
Ministero Istruzione Universita Ricerca for research grants.

\end{document}